\documentclass[pra,showpacs,nofootinbib]{revtex4-2}

\usepackage{hyperref}
\usepackage{enumitem}
\usepackage{amsmath}
\usepackage{amssymb}
\usepackage{amsthm}
\usepackage{graphics}
\usepackage{graphicx}
\usepackage{subcaption}
\usepackage{color}
\usepackage{mathrsfs}
\usepackage[usenames,dvipsnames,svgnames]{xcolor}
\usepackage{stmaryrd}
\usepackage{calc}
\usepackage{url}

\long\def\ca#1\cb{}

\def\bra#1{\langle#1|}

\def\inpd#1#2{\langle#1|#2\rangle }
\def\ket#1{|#1\rangle }

\def\Tr#1{\textrm{Tr}\left(#1\right)}

\def\locc{\overline{\textrm{LOCC}}}  
\def\mmod#1{(\bmod{~#1})}
\newcommand{\Mod}[1]{\ (\mathrm{mod}\ #1)}
\def\myeq#1{Eq.~\eqref{#1}}
\def\AC{{\cal A}}
\def\BC{{\cal B}}

\def\DC{{\cal D}}

\def\HC{{\cal H}}
\def\IC{{\cal I}}

\def\LC{{\cal L}}
\def\MC{{\cal M}}

\def\PC{{\cal P}}

\def\SC{{\cal S}}

\def\ZC{{\cal Z}}
\def\endproof{{\hspace{\stretch{1}}$\blacksquare$}}

\newcommand{\norm}[1]{\left\vert#1\right\vert} 
\newcommand{\normm}[1]{\left\|#1\right\|} 

\newtheorem{thm1}{Theorem}

\newtheorem{thm2}[thm1]{Theorem}
\newtheorem{thm3}[thm1]{Theorem}

\newtheorem{thm4}[thm1]{Theorem}
\newtheorem{thm5}[thm1]{Theorem}
\newtheorem{thm6}[thm1]{Theorem}
\newtheorem{thm7}[thm1]{Theorem}
\newtheorem{thm8}[thm1]{Theorem}

\newtheorem{lem1}{Lemma}
\newtheorem{cor1}{Corollary}

\newtheorem{lem2}[lem1]{Lemma}
\newtheorem{lem3}[lem1]{Lemma}

\newtheorem{lem14}[lem1]{Lemma}

\begin{document}
\title{Local approximation for perfect discrimination of quantum states}
\author{Scott M. Cohen}
\email{cohensm52@gmail.com}
\affiliation{Department of Physics, Portland State University, Portland Oregon, USA 97201}

\begin{abstract}
Quantum state discrimination involves identifying a given state out of a set of possible states. When the states are mutually orthogonal, perfect state discrimination is always possible using a global measurement. In the case of multipartite systems when the parties are constrained to use multiple rounds of local operations and classical communication (LOCC), perfect state discrimination is often impossible even with the use of \emph{asymptotic LOCC}, wherein an error is allowed but must vanish in the limit of an infinite number of rounds. Utilizing our recent results on asymptotic LOCC, we derive a lower bound on the error probability for LOCC discrimination of any given set of mutually orthogonal pure states. Informed by the insights gained from this lower bound, we are able to prove necessary conditions for perfect state discrimination by asymptotic LOCC. We then illustrate by example the power of these necessary conditions in significantly simplifying the determination of whether perfect discrimination of a given set of states can be accomplished arbitrarily closely using LOCC. The latter examples include a proof that perfect discrimination by asymptotic LOCC is impossible for a certain subset of \emph{minimal} unextendible product bases (UPB), where minimal means that for the given multipartite system, no UPB with a smaller number of states can exist. We also give a simple proof that what has been called \emph{strong nonlocality without entanglement} is considerably stronger than had previously been demonstrated.

\end{abstract}

\date{\today}
\maketitle

\section{Introduction}\label{sec1}
Nonlocality in quantum physics is a concept that has long intrigued researchers. Recognizing that this concept can mean different things in different contexts, Griffiths \cite{RBGLocality} has drawn a distinction between the use of the term nonlocality to describe the properties of quantum systems, on the one hand, and nonlocal influences between systems, on the other. The latter is a subject that continues to be widely debated, even to the present day.\footnote{While we have not studied this issue in depth, we do lean in the direction of Professor Griffiths' view that there is no evidence for these nonlocal influences.} Nonlocal properties of quantum systems are less controversial, but there are at times differing conceptions of what they entail. One such example, wherein a set of quantum states can exhibit nonlocal properties even when no one of those states is itself nonlocal, has received a great deal of attention in recent years. This property, first discovered over twenty years ago in the seminal work of \cite{Bennett9}, is commonly referred to as nonlocality without entanglement (NLWE) and arises in the context of quantum state discrimination \cite{Bergou,PeresWootters,ChitambarHsiehHierarchy,ChildsLeung,KKB}, wherein a party or parties are tasked with determining in which one of a known set of states their shared system had been prepared. Quantum state discrimination is a key paradigm in quantum information processing and quantum computing \cite{NielsenChuang}, and can also play a role in any experiment for which there exists enough \emph{a priori} information to narrow down the possible outcomes of that experiment.

When the set of states consists only of product states, having no entanglement \cite{HoroRMP} and therefore exhibiting no nonlocal properties individually, it may still turn out that taken as a set, the collection does indeed behave nonlocally. As shown in Ref.~\cite{Bennett9}, a particular set of mutually orthogonal product states on a $3\times3$ system cannot be perfectly discriminated when the parties are restricted to multiple rounds of measuring their local part of the system and communicating their outcomes to the other parties---a process known as LOCC, for local operations and classical communication---even though this can be easily accomplished by a single global measurement on the entire system taken as a whole. It is in this sense that the set of states exhibits nonlocality: when the system is measured locally it behaves differently than when it is measured globally.

The proof of NLWE given in Ref.~\cite{Bennett9} involved a long, complicated argument. The reason was that their aim was not simply to exclude the possibility of perfect local discrimination of the states, which is actually quite easily shown, but importantly, that the parties could not even approach accomplishing this task arbitrarily closely. We believe the latter definition of NLWE is the proper one to follow, and we will do so throughout this paper: A set of mutually orthogonal product states exhibits NLWE if and only if perfect discrimination of that set is impossible even when an error is allowed but must vanish in the limit of an infinite number of rounds. By not allowing for this vanishing error, one overlooks the fact that any measurement will be subject to experimental imperfections, that nothing is ever accomplished perfectly in the real world. As a consequence of these unavoidable imperfections, it is more appropriate to ask whether or not a task can be accomplished arbitrarily closely, and if not, the amount of error that is impossible to avoid. Many of us over the years have failed to clearly understand the important distinction that considering infinite-round protocols, alone, is not the same as allowing for vanishing error. In the former approach, it would be sufficient to show that no party can make a first local measurement without destroying orthogonality of the set (see, for example, Appendix B of \cite{myLDPE}): if no one can start the protocol, they cannot continue it indefinitely (or, at all). This approach fails to consider all possible \emph{sequences} of LOCC protocols of steadily increasing number of rounds, wherein as one proceeds through the sequence, the error incurred might become smaller and smaller, approaching zero asymptotically. To better understand what is missed, let us consider how one may think about these things.

LOCC protocols are commonly viewed as a tree graph with a single root node representing the situation before any party has measured. From the root node, the tree branches to multiple nodes, each one representing an outcome of the first measurement, which is local, being implemented by only one party. From each of these nodes, the tree continues to branch to more nodes, each set of \emph{child} nodes of a given \emph{parent} node representing outcomes of the local measurement performed at that stage of the protocol. A finite branch of the tree starts at the root and continues until it reaches a node that has no children, denoted as a \emph{leaf} node. In the limit of an infinite number of rounds, there may be branches that never terminate and are of infinite length. As shown in \cite{myProdPaths}, each branch (finite or infinite) corresponds directly to a continuous path of product operators through a particular subset of operator space. Significantly, for a given protocol, each such path is \emph{piecewise local}, which means that it consists of straight line segments along which the product operator changes only in one party's local part. For example on a bipartite system, this might be represented as $\left[(1-x)\AC+x\AC^\prime\right]\otimes\BC$, which is a line stretching from $\AC\otimes\BC$ to $\AC^\prime\otimes\BC$ as $x$ ranges from $0$ to $1$, $\AC^\prime$ corresponding to one outcome of a local measurement by Alice, $\AC$ being the cumulative effect of Alice's actions up to (and preceding) this latest measurement. Only the $A$-part changes, the $B$-part remains unchanged. This piecewise local property applies even to infinite branches, which then consist of many (an infinite number of) infinitesimally short pieces.

To understand why only considering individual infinite-round protocols overlooks possibilities, let us recall how we learned about integrals in our introductory calculus classes: any curve can be approximated arbitrarily closely by a piecewise constant curve, and this provides a way to approximate the area under the original curve. In the limit that the number of constant pieces goes to infinity, the piecewise constant curve asymptotically approaches the original curve, which in general is \emph{not} piecewise constant. Similarly, there exist sequences of LOCC protocols for which each protocol in the sequence corresponds to piecewise local paths in operator space, but for which the limit of this sequence corresponds to paths which are \emph{not} piecewise local. In particular, it may well be that the limit of a sequence of LOCC protocols corresponds to an initial measurement that is not local, and such sequences are \emph{not} excluded by simply demonstrating that the only initial \emph{local} measurement that does not destroy orthogonality of the original set is a measurement for which all outcomes are proportional to the identity operator (a \emph{trivial} measurement). Instead, as we will see below, it is sufficient to show that any non-trivial initial \emph{separable} measurement operator that is arbitrarily close to $I_\HC$ destroys orthogonality, see Corollary~\ref{cor1}.

In an effort to ensure these ideas are clear, let us divide the class of infinite-round protocols into two distinct subclasses \cite{WinterLeung}. The first subclass involves sequences of protocols where each subsequent protocol in a given sequence differs from the preceding protocol simply by adding more rounds of communication, but without changing the local measurements implemented in earlier rounds. Since the earlier rounds are unchanged, the branches remain piecewise local even in the infinite limit. Thus, one obtains a valid LOCC protocol in this limit, albeit one having an infinite number of rounds, so this subclass may be seen as being a part of LOCC. The second subclass includes limits of sequences in which measurements made at the earlier rounds are changed from one protocol in the sequence to the next. This subclass must be included to obtain the asymptotic LOCC discussed above, and its inclusion gives rise to the topological closure of LOCC, which we denote as $\overline{\textrm{LOCC}}$ in the sequel. By changing those earlier rounds, the branches need not correspond to piecewise local paths in the infinite limit (see the comparison to limits of piecewise constant curves in the preceding paragraph) and as such, in this limit, one may fail to obtain a valid LOCC protocol.

Let us review the main result, Theorem~$1$, of Ref.~\cite{myProdPaths}, where we consider a measurement as a positive operator valued measure (POVM) consisting of a set of positive semidefinite operators, $E_j$, known as POVM elements. Note that $\MC\in\locc$ means there exists a sequence of LOCC protocols, the $n$th such protocol implementing measurement $\MC_n$, such that $\lim_{n\to\infty}\MC_n=\MC$.

{\bf Theorem $1$} of \cite{myProdPaths}. \emph{If ${\cal M}\in\overline{\textrm{LOCC}}$, with measurement ${\cal M}$ consisting of POVM elements $E_j$, then for each $j$, there exists a continuous, monotonic path of product operators from ${\cal I}_{\cal H}$ to a point on the (half-open) line segment $(0,E_j]$, and this path lies entirely within the geometric object, $\ZC_{\cal M}=\sum_j[0,E_j]$, which is known as a \emph{zonotope}.}

\noindent An alternative, but equivalent, definition of the zonotope just introduced is $\ZC_\MC:=\left\{z\left\vert z=\sum_{j}c_{j}E_j,~0\le c_j\le1~\forall{j}\right.\right\}$. In addition, by \emph{monotonic}, we mean that the trace of the product operators is non-increasing along these paths. These paths, which need not themselves be piecewise local, are found as the limit of a sequence of piecewise local paths, the latter being associated with that sequence of LOCC protocols, the limit of which implements $\MC$. Of course, without the restriction that the paths lie within $\ZC_\MC$, these paths would always exist. That is, there are always such paths between any pair of product operators. For example, one path from $\AC\otimes\BC$ to $\AC^\prime\otimes\BC^\prime$ would be $[(1-x)\AC+x\AC^\prime]\otimes\BC$ followed by $\AC^\prime\otimes[(1-y)\BC+y\BC^\prime]$. As is amply illustrated by the examples in Ref.~\cite{myProdPaths}, however, there are many measurements for which there are no paths of product operators starting at $I_\HC$ and lying within $\ZC_\MC$. It is worth noting that those examples were drawn from well-studied cases of local state discrimination, for which much of the work in \cite{myProdPaths} involved determining the most general separable measurement $\MC$ capable of perfectly discriminating the given set, and then showing that the requisite paths of product operators within $\ZC_\MC$ do not exist. Here, we simplify things by finding ways to reach these conclusions for given sets of states without the need to know anything about what measurements can accomplish the task successfully. Another observation is that since the paths considered here are continuous and starting at $I_\HC$, they require the existence of a positive semidefinite product operator lying within $\ZC_\MC$ at every distance, $R$, from $I_\HC$ in the range $0\le R\le d(I_\HC,E_j)$, with $d(X,Y)$ the distance between (normalized operators) $X$ and $Y$. Suppose such continuous paths do not exist. Then it seems a reasonable guess that the error incurred by any LOCC protocol used for implementing the desired POVM will be in some way related to how far away from $\ZC_\MC$ one must stray in order to find such paths, and in an attempt to lower bound this error, one may then consider, for each $R$, how far it is from $\ZC_\MC$ to the nearest positive semidefinite product operator. This is part of the motivation for the present work, in which we prove a result that is similar, but not quite identical, to what we have just conjectured. Our result differs from the ideas just described in one very important aspect: in order to know the distance of an operator from $\ZC_\MC$, one needs to know the measurement, $\MC$. It turns out that knowledge of what measurement(s) might be successful is not needed, all we need to know is the set of states one is setting out to discriminate.

The remainder of the paper is organized as follows: In Section~\ref{sec2}, we use the insights of \cite{myProdPaths} to derive a lower bound on the probability of error, $p_\textrm{err}$, in locally discriminating any mutually orthogonal set of pure states. We allow for limits of sequences of protocols, discussed above, going beyond LOCC itself to include $\locc$. These arguments demonstrate that $p_\textrm{err}>0$ implies the set of states cannot be perfectly discriminated within $\locc$, which would require asymptotically vanishing error. Unfortunately, we have found this lower bound to be difficult to compute. Nonetheless, in Section~\ref{sec3}, we use the insights gleaned from this lower bound to prove two theorems providing necessary conditions that a given set of states can be perfectly discriminated within $\locc$, and then in Section~\ref{sec4}, we give examples where these theorems easily demonstrate that this is impossible. We also show that perfect discrimination within $\locc$ is impossible when the set of states is an unextendible product basis with the minimal number of states for the associated multipartite Hilbert space, and when in addition, the local dimensions are such that the number of states satisfies $N\ge2(d_\alpha-1)+1$. Finally, we end with our conclusions.

\section{Error probability for discriminating any set of orthogonal pure states by $\overline{\textrm{LOCC}}$}\label{sec2}
In this section, we begin by considering the error incurred in using LOCC to discriminate a set $\SC$ of $N$ orthogonal pure states on Hilbert space $\HC$: $\SC=\{\eta_m,\ket{\Psi_m}\}$, given with \emph{a priori} probabilities $\eta_m>0$, $\sum_m\eta_m=1$, and $\inpd{\Psi_m}{\Psi_n}=\delta_{mn}$. For an arbitrary set of orthogonal states $\ket{\Psi_m}$, not necessarily a complete basis of the Hilbert space, there will generally be a number of possible global measurements that perfectly discriminate those states. In general, however, when these states describe a multipartite system, there may be constraints on the actions the parties are able to perform, and under such circumstances, any given global measurement, $\MC_g$, may be impossible. Instead, the parties may be restricted to using LOCC in their efforts to discriminate the state, and they may be forced to utilize a different measurement, say
\begin{align}\label{eqn20}
	\MC_Q=\left\{Q_i\left\vert\sum_iQ_i=I_\HC,Q_i\ge0\right.\right\}.
\end{align}
It may be that $\MC_Q$ can be implemented by LOCC, at least arbitrarily closely: $\MC_Q\in\overline{\textrm{LOCC}}$. If not, then the question arises, how well can the parties do in discriminating the state if they are able to perform the best possible LOCC measurement, $\MC_Q$?

Define $P_R$ to be the set of positive semidefinite product operators acting on $\HC$ and lying at a distance $R$ from the identity operator $I_\HC$ (distances \emph{between normalized operators} are measured using the Frobenius norm, defined above Eq.~\eqref{eqn1105}, below), and also define
\begin{align}\label{eqn120}
	\Pi=\sum_m\sqrt{\eta_m}\Psi_m,
\end{align}
with $\Psi_m=\ket{\Psi_m}\bra{\Psi_m}$. Then, in Appendix~\ref{App6}, we prove that
\begin{thm1}\label{thm1}
	Given any set of mutually orthogonal pure states, the probability of error for local state discrimination of this set is lower-bounded as
	\begin{align}\label{eqn1104}
		p_{\textrm{err}}&\ge\frac{1}{2}\max_R\min_{\substack{Q\in\PC_R\\z\in\ZC_\Psi}}\normm{\frac{\Pi Q\Pi-z}{\Tr{\Pi^2Q}}}^2,
	\end{align}
\end{thm1}
\noindent where $\normm{\cdot}$ is the Frobenius norm, and $\max_R$ is taken over the range $0\le R\le\sqrt{(D-1)/D}$, see Appendix~\ref{App6} for details. With $\ZC_\Psi:=\left\{z\left\vert z=\sum_{m}c_m\Psi_m,~0\le c_m\le1~\forall{m}\right.\right\}$, it is straightforward to see that the minimum over $z\in\ZC_\Psi$ is achieved at $z=\sum_m\eta_m\bra{\Psi_m}Q\ket{\Psi_m}\Psi_m$. Note that the definition of $\ZC_\Psi$ closely adheres to how $\ZC_\MC$ was defined above for a measurement $\MC$, even though the set of operators, $\Psi_m$, does not in general constitute a complete measurement. When $p_{\textrm{err}}=0$ and the set of states can be perfectly discriminated using $\locc$, then operator $\Pi$ effectively takes the paths of Ref.~\cite{myProdPaths}, which are associated with an actual complete measurement $\MC$ and lie within $\ZC_\MC$, projecting (and scaling, by the $\eta_m$) them into $\ZC_\Psi$. Thus, we obtain continuous paths of product operators that lie entirely within $\ZC_\Psi$, and this is why we need not know $\ZC_\MC$ or the optimal measurement, $\MC$, that is to be used. Given this observation, it would seem to make sense to consider measurements $\MC$ such that $\ZC_\Psi\subseteq\ZC_\MC$ whenever possible, although it isn't entirely clear this would necessarily minimize the error.

Calculating this lower bound on $p_{\textrm{err}}$ appears to be extremely challenging in practice. Approaching this problem analytically is prohibitively difficult except for the smallest systems; that is, for two qubits, in which case it is \emph{merely} very challenging. For the latter case, we have been able to show for discriminating the four Bell states \cite{NielsenChuang} by LOCC---when they are given with equal \emph{a priori} probabilities, $\eta_m=1/4$---that our lower bound is $p_{\textrm{err}}=1/4$, which is just a factor of two smaller than the known optimal strategy \cite{WatrousYu2015}. That this is the correct order of magnitude may be seen as an encouraging sign. Calculating this lower bound does not appear to fall into any of the classes that admit an efficient numerical approach, however. Therefore, one would need access to significant computational resources to obtain a result with a high degree of confidence it is truly a lower bound. Therefore in the next section, we will obtain powerful necessary conditions for the possibility of perfect state discrimination by $\locc$ of any given set of mutually orthogonal pure states. Note that the \emph{a priori} probabilities, $\eta_m>0$, are only relevant to the question of the amount of error incurred and not to whether or not perfect discrimination is possible.

\section{Necessary conditions for perfect state discrimination by $\locc$}\label{sec3}
Our first necessary condition is obtained as follows. If there exists R such that $\Delta_R>0$, then $p_{\textrm err}>0$---or alternatively (recalling the perspective of the result of Ref.~\cite{myProdPaths}), the required continuous paths of product operators whose projection by $\Pi$ lies entirely within $\ZC_\Psi$ do not exist---then perfect discrimination of the set of states by $\locc$ is impossible. That is,
\begin{thm3}\label{thm3}
	Given a set of mutually orthogonal quantum states, $\{\ket{\Psi_m}\}$, if for any fixed state $\ket{\Psi_n}$, no continuous path of positive semidefinite product operators, say $Q_i(s)$, exists such that the following two conditions hold:
	\begin{enumerate}
		\item the path begins at $I_\HC$ and ends at some fixed positive semidefinite product operator, $Q_i$, where $\Pi Q_i\Pi\propto\Psi_n$, and index $i$ will generally depend on index $n$;
		\item for every $s$, $Q_i(s)$ is diagonal in the (partial) basis of the $\ket{\Psi_m}$;
	\end{enumerate} 
	then this set of states cannot be perfectly discriminated within $\locc$. Note that the condition that $Q_i(s)$ is diagonal in the $\ket{\Psi_m}$ is equivalent to $\Pi Q_i(s)\Pi$ lying within $\ZC_\Psi$.
\end{thm3}
\proof Suppose there exists measurement $\MC_Q\in\locc$ as in \myeq{eqn20} that perfectly discriminates the given set of states. Then by Theorem~$1$ of Ref.~\cite{myProdPaths}, for each $Q_i\in\MC$, there exists a continuous path of product operators $Q_i(s)$ extending from $I_\HC$ to $Q_i$ and lying entirely within $\ZC_{\MC}$. Furthermore, each $Q_i\ge0$ identifies without error one of the states in the set, say $\Psi_n$, or in other words, $\Tr{Q_i\Psi_m}=\delta_{mn}q_{in}$, which since $\Psi_n\ge0$ as well, means that $Q_i\Psi_m=0=\Psi_mQ_i$ for all $m\ne n$. This implies that $\Pi Q_i\Pi=q_{in}\Psi_n$, for the given fixed $n$, with $q_{in}\ge0$. Now, from the proof of Theorem~\ref{thm1} in Ref.~\cite{myProdPaths}, we know that $Q_i(s)=\sum_jc_{ij}(s)Q_j$ with $c_{ij}(s)\ge0$. This leads to $\Pi Q_i(s)\Pi=\sum_jc_{ij}(s)\Pi Q_j\Pi=\sum_j\sum_mq_{jm}c_{ij}(s)\Psi_m\in\ZC_\Psi$. Since the path terminates at $Q_i$, and as we've seen, $\Pi Q_i\Pi\propto\Psi_n$ for some $n$, the proof of this theorem is complete.\endproof

\noindent We will use this Theorem in the next section to prove that an unextendible product basis consisting of the minimal number of states cannot be perfectly discriminated using $\locc$ if the local dimensions are such that the number of states satisfies $N\ge2(d_\alpha-1)+1$. 

Noting that we can write our positive semidefinite path of operators as $Q(s)=K(s)^\dag K(s)$, the condition in Theorem~\ref{thm3} that $Q(s)$ is diagonal in the partial basis of the $\ket{\Psi_m}$ is equivalent to orthogonality of the new states, $K(s)\ket{\Psi_m}$. Thus, we have the following corollary to Theorem~\ref{thm3}.
\begin{cor1}\label{cor1}
	If a given set of mutually orthogonal quantum states, $\ket{\Psi_m}$, can be perfectly discriminated within $\locc$, then there exists a continuous path of product operators, $K(s)$, such that for every $s$, the states $K(s)\ket{\Psi_m}$ remain orthogonal along the entire path.
\end{cor1}
\noindent Notice how this generalizes the observation, discussed here in the Introduction, that the initial, local measurement in any LOCC protocol must preserve orthogonality of the states. Here we see instead that for $\locc$, orthogonality is preserved along entire continuous paths of operators, and also that while $K(s)$ must be a product (separable measurement) operator, this path need not be piecewise local, so need not correspond to a series of local measurements.

The proof of the next result will use an extension of the notion, introduced in \cite{IBM_CMP}, of partitions of the states of an unextendible product basis (UPB) amongst the various parties. Note, however, that this theorem is general, being applicable to any set of states, not just UPBs. Let us review these ideas before proceeding to the theorem, itself.

An UPB is a set, $\SC$, of $N$ mutually orthogonal product states on multipartite Hilbert space $\HC$ such that there is no other product state on $\HC$ that is orthogonal to all the original $N$ states in the UPB. In principle, a complete product basis of $\HC$ is unextendible, but one is usually only concerned with \emph{partial} bases, such that the $N$ states do not span the complete space $\HC$. The following lemma was proved in \cite{IBM_CMP}.
\begin{lem14}\cite{IBM_CMP}\label{lem14}
	Let $\pi$ be a partition of $\SC$ into $P$ disjoint subsets equal to the number of parties: $\SC = S_1\cup S_2\cup\cdots\cup S_P$. Let
	$r_\alpha = \textrm{rank}\{\ket{\psi_j^{(\alpha)}}: \ket{\Psi_j}\in S_\alpha \}$ be the local rank of subset $S_\alpha$ as seen by the $\alpha$th party. Then $\SC$ is extendible if and only if there exists a partition $\pi$ such that for all $\alpha = 1,\ldots, P$, the local rank of the $\alpha$th subset is less than the dimensionality of the $\alpha$th party’s Hilbert space. That is to say, $\SC$ is extendible if and only if there exists $\pi$ such that for all $\alpha$, $r_\alpha < d_\alpha$.
\end{lem14}
\noindent The partitioning introduced in this lemma can be understood as a way of distributing ``the job of being orthogonal to a new product state" \cite{IBM_PRL} among the various parties. If for every such partition, at least one party's local states---say party $\alpha$ with set of local states $S_\alpha$---span the full local Hilbert space, then there is no state orthogonal to all the states in $S_\alpha$, and party $\alpha$ fails to fulfill its role of being orthogonal to an additional product state. If for every partition at least one party fails in this role, then there is no additional product state orthogonal to all the states in $\SC$. In other words, under these circumstances, the original set is unextendible.

We are now ready to prove our second necessary condition for perfect state discrimination by $\locc$.
\begin{thm4}\label{thm4}
	Consider any mutually orthogonal set of product states $\SC=\left\{\ket{\Psi_j}=\bigotimes_{\alpha}\ket{\psi_j^{(\alpha)}}\right\}$. For each party $\alpha$, define the subset of all index pairs, $J_\alpha=\left\{(i,j)\left\vert\inpd{\psi_i^{(\alpha)}}{\psi_j^{(\alpha)}}=0;~\inpd{\psi_i^{(\beta)}}{\psi_j^{(\beta)}}\ne0~\forall{\beta\ne\alpha}\right.\right\}$. If for every party $\alpha$ the set of dyads, $\left\{\ket{\psi_i^{(\alpha)}}\bra{\psi_j^{(\alpha)}}\right\}_{(i,j)\in J_\alpha}$, spans a space of dimension $d_\alpha^2-1$, then this set of product states cannot be perfectly discriminated within $\locc$.
\end{thm4}
\noindent Throughout the remainder of this paper, we will refer to kets $\ket{\psi_j^{(\alpha)}}$ as the \emph{local states} on $\HC_\alpha$. The idea of the proof is that when these dyads span a space of dimension $d_\alpha^2-1$, there is one and only one operator orthogonal to all of them, that being the identity operator, $I_\alpha$. If this is true for all parties, there can be no global product operator that is orthogonal to all these global dyads and is close to but not proportional to $I_\HC$, and then by Theorem~\ref{thm1}, perfect discrimination by $\locc$ is impossible.

\proof We extend the notion of partitioning to the set of dyads $\DC=\{\ket{\Psi_i}\bra{\Psi_j}\}_{j\ne i}$. Let $\hat\pi$ be such a partition, yielding $\DC=D_1\cup D_2\cup\cdots\cup D_P$, which we will understand as a way to distribute the ``job of being orthogonal" to a product operator $Q=\bigotimes_\alpha Q^{(\alpha)}$. That is, given $\hat\pi(s)$, $\Tr{Q^{(\alpha)}(s)\ket{\psi_i^{(\alpha)}}\bra{\psi_j^{(\alpha)}}}=0$ for all $\ket{\Psi_i}\bra{\Psi_j}\in D_\alpha$.

By Theorem~\ref{thm3}, if the states of $\SC$ can be perfectly discriminated within $\locc$, then there exists a continuous path of positive semidefinite product operators $Q(s)$ such that for every $s$, $Q(s)$ is diagonal in the partial basis of the states in $\SC$: $\bra{\Psi_i}Q(s)\ket{\Psi_j}=\delta_{ij}\bra{\Psi_i}Q(s)\ket{\Psi_i}$. Now, for each distinct $s$, one may assign a different partition $\hat\pi(s)$ to distribute the orthogonality job. However, given that there is a finite number of states, $N$, there is also a finite number of dyads, $N(N-1)$, and thus there is a finite number of distinct partitions that can be used here. If for any given partition, each (and every) party $\alpha$ is given a set of dyads spanning a subspace of dimension $d_\alpha^2-1$, then for that partition there is one and only one operator $Q^{(\alpha)}$ orthogonal to all of that party's dyads, and thus there is one and only one operator $Q$ orthogonal to all the multipartite dyads, $\ket{\Psi_i}\bra{\Psi_j}$ for $j\ne i$. Given there are a finite number of partitions, there are then only a finite number of operators orthogonal to all the multipartite dyads, and there cannot be a continuous path of operators from $I_\HC$ to anywhere. Indeed, given that the dyads have been distributed according to the index sets $J_\alpha$, each of these local dyads is traceless, and thus orthogonal to the identity operator, $I_\alpha$. Thus, for each such partition, the only operator orthogonal to these dyads is $Q(s)=I_\HC$, which is a \emph{point} and not a \emph{path} that leaves $I_\HC$, as is required.

There are two points that need clarification here. First, partitioning dyads according to $J_\alpha$ omits dyads, which are therefore not given to any of the parties. As already noted elsewhere, this is not an issue for our proof because including those additional dyads can only \emph{increase} the space spanned by the dyads given to any given party, so can only further constrain operators $Q^{(\alpha)}(s)$ orthogonal to these local dyads. The second point is that there are many partitions that do not conform to $J_\alpha$. For example, there will generally be partitions such that dyad $\ket{\Psi_i}\bra{\Psi_j}$ is given to party $\alpha$ even when the corresponding local states on $\alpha$ are not themselves orthogonal. As explained in the next paragraph, we will not need to consider any of these other partitions.

The reason we can restrict consideration to those partitions that follow $J_\alpha$ is that these are the only ones relevant for small enough $s$. Let us see why this is so. Since this path of operators starts at $Q(0)=I_\HC$, then by continuity, there exists $Q(s)$ for small enough $s$ which is arbitrarily close to $I_\HC$. If $\inpd{\psi_i^{(\beta)}}{\psi_j^{(\beta)}}\ne0$, then for small enough $s$, $\bra{\psi_i^{(\beta)}}Q^{(\beta)}(s)\ket{\psi_j^{(\beta)}}$ is also non-vanishing, and $\ket{\psi_i^{(\beta)}}\bra{\psi_j^{(\beta)}}$ is not orthogonal to $Q^{(\beta)}(s)$. To see this formally, one may measure distances between operators on $\HC_\beta$ by the Frobenius norm, $\normm{X}=\sqrt{\sum_{k,l}\norm{X_{kl}}^2}\ge\norm{\bra{\psi_i^{(\beta)}}X\ket{\psi_j^{(\beta)}}}$ for some fixed $i,j$ (no sum).\footnote{This inequality is obvious when $\inpd{\psi_i^{(\beta)}}{\psi_j^{(\beta)}}=0$, and it is straightforward to show that $\normm{X}\ge\norm{\bra{\psi_i^{(\beta)}}X\ket{\psi_j^{(\beta)}}}$ also holds for any non-orthogonal pair of states.} Then, for $\norm{\inpd{\psi_i^{(\beta)}}{\psi_j^{(\beta)}}}=r\gg\epsilon>0$ and $Q^{(\beta)}(s)$ within $\epsilon$ of $I_\beta$, we have
\begin{align}\label{eqn1105}
	\epsilon>\normm{I_\beta-Q^{(\beta)}(s)}\ge\norm{\bra{\psi_i^{(\beta)}}(I_\beta-Q^{(\beta)}(s))\ket{\psi_j^{(\beta)}}}=\norm{r-\bra{\psi_i^{(\beta)}}Q^{(\beta)}(s)\ket{\psi_j^{(\beta)}}},
\end{align}
implying that $\norm{\bra{\psi_i^{(\beta)}}Q^{(\beta)}\ket{\psi_j^{(\beta)}}}\approx r\gg\epsilon>0$ and $Q^{(\beta)}(s)$ is not orthogonal to the corresponding dyad, $\ket{\psi_j^{(\beta)}}\bra{\psi_i^{(\beta)}}$.
If $\inpd{\psi_i^{(\beta)}}{\psi_j^{(\beta)}}\ne0$ for all $\beta\ne\alpha$, then the job of $\ket{\Psi_j}\bra{\Psi_i}$ being orthogonal to $Q(s)$ for small enough $s$ must be assigned to party $\alpha$, and this completes the proof. 
\endproof

As a simple illustration of how this works, consider the set of two-qubit states, which can be perfectly discriminated by LOCC, $\SC=\{\ket{0}\ket{0},\ket{0}\ket{1},\ket{1}\ket{+},\ket{1}\ket{-}\}$, with $\ket{\pm}=(\ket{0}\pm\ket{1})/\sqrt{2}$. For the second party, we have the index set $J_2=\{(1,2),(2,1),(3,4),(4,3)\}$. The corresponding dyads are $\{\ket{0}\bra{1},\ket{1}\bra{0},\ket{+}\bra{-},\ket{-}\bra{+}\}$. This set spans a space of dimension $d_2^2-1=3$, indicating there is a unique local operator orthogonal to the entire set, that being the identity operator $I_2$. Thus, there is no product operator $A\otimes B$ close to $I_\HC$ that \emph{does not destroy orthogonality} of the original set $\SC$, except possibly those with $B\propto I_2$. Looking at the first party, $J_1=\{(1,3),(2,3),(1,4),(2,4),(3,1),(3,2),(4,1),(4,2)\}$, with corresponding dyads, $\{\ket{0}\bra{1},\ket{1}\bra{0}\}$, which span a space of dimension only $2<d_1^2-1$. This leaves party $1$ with a range of possible measurement operators close to the identity (anything diagonal in the standard basis is acceptable) and as is fairly obvious, this party can indeed initiate a successful LOCC protocol.

If the local parts of dyads $\ket{\Psi_m}\bra{\Psi_n},~m\ne n$, are to span a subspace of dimension $d_\alpha^2-1$ for each party $\alpha$, there must be enough pairs of states in the original set to distribute to all parties, $N(N-1)\ge\sum_\alpha(d_\alpha^2-1)=:T$. This provides a lower bound on the number of states, $N\ge\left\lceil\frac{1}{2}+\sqrt{T+\frac{1}{4}}~\right\rceil$, which is smaller than the minimal number of states in a UPB on the same system. However, we do not know if there exist sets of states that achieve this new lower bound while still exhibiting NLWE, or if a larger number is needed.

Additional examples illustrating the power of these ideas will be found in the next section.

\section{Applications}\label{sec4}
In this section, we illustrate the results of the preceding one with a few explicit examples.
\subsection{The Rotated Domino States and the Tiles UPB}\label{ssec1}
It is perhaps worth showing how easy it can sometimes be to prove that certain sets of states cannot be perfectly discriminated within $\locc$. Let us begin with two well-known sets of states for which this has previously been proven \cite{ChildsLeung,myProdPaths} using more---sometimes, much, much more---complicated arguments. The rotated domino states are \cite{Bennett9}
\begin{align}\label{eqn205}
	\ket{\Psi_1}&=\ket{1}\otimes\ket{1}\notag\\
	\ket{\Psi_2}&=\ket{0}\otimes(\cos{\theta_1}\ket{0}+\sin{\theta_1}\ket{1})\notag\\
	\ket{\Psi_3}&=\ket{0}\otimes(\sin{\theta_1}\ket{0}-\cos{\theta_1}\ket{1})\notag\\
	\ket{\Psi_4}&=(\cos{\theta_2}\ket{0}+\sin{\theta_2}\ket{1})\otimes\ket{2}\notag\\
	\ket{\Psi_5}&=(\sin{\theta_2}\ket{0}-\cos{\theta_2}\ket{1})\otimes\ket{2}\notag\\
	\ket{\Psi_6}&=\ket{2}\otimes(\cos{\theta_3}\ket{1}+\sin{\theta_3}\ket{2})\notag\\
	\ket{\Psi_7}&=\ket{2}\otimes(\sin{\theta_3}\ket{1}-\cos{\theta_3}\ket{2})\notag\\
	\ket{\Psi_8}&=(\cos{\theta_4}\ket{1}+\sin{\theta_4}\ket{2})\otimes\ket{0}\notag\\
	\ket{\Psi_9}&=(\sin{\theta_4}\ket{1}-\cos{\theta_4}\ket{2})\otimes\ket{0}
\end{align}
with $0<\theta_j\le\pi/4$. We can easily show these states cannot be perfectly discriminated using $\locc$.
\begin{thm5}\label{thm5}
	The rotated domino states of \myeq{eqn205} cannot be perfectly discriminated using $\locc$.
\end{thm5}
\proof We will use Theorem~\ref{thm4}, so identify for the first party, $J_1\supset\{(3,9),(5,7),(3,7),(4,5),(8,9)\}$, corresponding to dyads,
\begin{align}\label{eqn206}
	\ket{0}(\sin{\theta_4}\bra{1}-\cos{\theta_4}\bra{2})&&(\sin{\theta_4}\ket{1}-\cos{\theta_4}\ket{2})\bra{0}\notag\\
	\ket{2}(\sin{\theta_2}\bra{0}-\cos{\theta_2}\bra{1})&&(\sin{\theta_2}\ket{0}-\cos{\theta_2}\ket{1})\bra{2}\notag\\
	\ket{0}(\bra{2}&&(\ket{2}\bra{0}\notag\\
	(\cos{\theta_2}\ket{0}+\sin{\theta_2}\ket{1})(\sin{\theta_2}\bra{0}-\cos{\theta_2}\bra{1})&&(\cos{\theta_4}\ket{1}+\sin{\theta_4}\ket{2})(\sin{\theta_4}\bra{1}-\cos{\theta_4}\bra{2}),	
\end{align}
and the Hermitian conjugates of the last pair of dyads are omitted, as they are not needed. To readily show these are linearly independent, consider
\begin{align}\label{eqn207}
	0=&c_1\ket{0}(\sin{\theta_4}\bra{1}-\cos{\theta_4}\bra{2})+c_2(\sin{\theta_4}\ket{1}-\cos{\theta_4}\ket{2})\bra{0}
	+c_3\ket{2}(\sin{\theta_2}\bra{0}-\cos{\theta_2}\bra{1})\notag\\
	&+c_4(\sin{\theta_2}\ket{0}-\cos{\theta_2}\ket{1})\bra{2}+c_5\ket{0}(\bra{2}+c_6\ket{2}\bra{0}+c_7(\cos{\theta_2}\ket{0}+\sin{\theta_2}\ket{1})(\sin{\theta_2}\bra{0}-\cos{\theta_2}\bra{1})\notag\\
	&+c_8(\cos{\theta_4}\ket{1}+\sin{\theta_4}\ket{2})(\sin{\theta_4}\bra{1}-\cos{\theta_4}\bra{2}).
\end{align}
It is very easy to show that this is satisfied if and only if all the coefficients vanish. The $0,0$ matrix element of \myeq{eqn207} gives $c_7=0$ and the $2,2$ element gives $c_8=0$. Then, each off-diagonal element shows that one of the remaining $c_j$ vanishes, and this encompasses all of them. Thus, $c_j=0$ for all $j$, and these $8$ dyads are linearly independent. Since there is a symmetry between the parties, then by Theorem~\ref{thm4}, this completes the proof.\endproof

\noindent Notice that while $\ket{\Psi_1}$ is needed to make this set a full basis, it does not appear in any of the dyads of \myeq{eqn206}. Therefore, the set still cannot be perfectly discriminated within $\locc$ even if this state is omitted.

Next, consider the states of the Tiles UPB, which is a subset of the Dominoes (unrotated, all $\theta_j=\pi/4$), except with $\ket{\Psi_1}$ replaced by $\ket{F}$,
\begin{align}\label{eqn301}
	\ket{F}&=\frac{1}{3}(\ket{0}+\ket{1}+\ket{2})\otimes(\ket{0}+\ket{1}+\ket{2})\notag\\
	\ket{\Psi_{3}}&=\frac{1}{\sqrt{2}}\ket{0}\otimes(\ket{0}-\ket{1})\notag\\
	\ket{\Psi_{5}}&=\frac{1}{\sqrt{2}}(\ket{0}-\ket{1})\otimes\ket{2}\notag\\
	\ket{\Psi_{7}}&=\frac{1}{\sqrt{2}}\ket{2}\otimes(\ket{1}-\ket{2})\notag\\
	\ket{\Psi_{9}}&=\frac{1}{\sqrt{2}}(\ket{1}-\ket{2})\otimes\ket{0}.
\end{align}
We have,
\begin{thm6}\label{thm6}
	The Tiles UPB of \myeq{eqn301} cannot be perfectly discriminated using $\locc$.
\end{thm6}
\proof In this case, we can use six of the same dyads as were used for the (rotated) dominoes, the ones in the first three rows of \myeq{eqn206} (but with $\theta_j=\pi/4$, as noted above). Then, instead of those in the fourth row there, include the local (on the first party) parts of $\ket{\Psi_5}\bra{F}$ and $\ket{\Psi_9}\bra{F}$. By following the same argument as was just used in the proof of the preceding theorem, it is easily seen that these are eight linearly independent dyads. Since there is again a symmetry between the parties, the proof is complete.\endproof

\subsection{``Strong" quantum nonlocality without entanglement}\label{ssec2}
We now turn to the results of Ref.~\cite{HalderPRL} concerning what they have denoted as strong nonlocality without entanglement. We have argued in the Introduction that these results are perhaps not as strong as claimed, or at least, as one might wish them to be. The authors of that paper have not proved that the sets of states discussed in their paper exhibit NLWE, according to how we believe NLWE should be understood, and therefore they have also not proved that they exhibit a stronger version of NLWE, as is their claim. Here, we show that for the first of their sets of states (on a tripartite system), their claims are nonetheless correct, that this set does exhibit NLWE, and we also show that it demonstrates NLWE across all bipartite cuts, therefore also exhibiting the stronger version of NLWE. 

Defining $\ket{j\pm k}=(\ket{j}\pm\ket{k})/\sqrt{2}$, the set of states on a $3\times3\times3$ system, given in Eq.~($4$) of Ref.~\cite{HalderPRL} as an example of what they call strong nonlocality without entanglement, is
\begin{align}\label{eqn201}
\ket{\Psi_{1\pm}}&=\ket{1}\ket{2}\ket{1\pm2}&\ket{\Psi_{4\pm}}&=\ket{1}\ket{3}\ket{1\pm3}\notag\\
\ket{\Psi_{7\pm}}&=\ket{2}\ket{3}\ket{1\pm2}&\ket{\Psi_{10\pm}}&=\ket{3}\ket{2}\ket{1\pm3}
\end{align}
and cyclic permutations of the local states in \myeq{eqn201}---so that for example, $\ket{\Psi_{2\pm}}=\ket{2}\ket{1\pm2}\ket{1}$, and generally $\ket{\Psi_{3j+k,\pm}}$ is obtained by permuting the local states in $\ket{\Psi_{3j+1,\pm}}$ $k-1$ times---along with $\ket{i}\ket{i}\ket{i}~i=1,2,3$.
We now demonstrate that this set of states does indeed exhibit strong NLWE, according to our definition. First, we show that this set exhibits NLWE.
\begin{thm7}\label{thm7}
	The set of states in Eq.~($4$) of \cite{HalderPRL} cannot be perfectly discriminated within $\locc$.
\end{thm7}
\proof We wish to apply Theorem~\ref{thm4}. Toward that end, we seek orthogonal pairs of local states on the first party, such that the corresponding pairs on the other parties are not orthogonal. By simple inspection, one easily finds there are many index pairs which satisfy this condition. We only need to find enough dyads to span a subspace of dimension $d_\alpha^2-1$; including more index pairs means more dyads, which cannot shrink the subspace that they span. Select index pairs, $(1+,2+),(1+,10+),(2+,5+),(3+,3-)$, and $(6+,6-)$, which lead to $10$ distinct dyads (including Hermitian conjugates) that (as explained in the proof of Theorem~\ref{thm4}) must be given to the first party, those dyads being $\ket{i}\bra{j}$ for all $i\ne j$ and $\ket{1+i}\bra{1-i},~i=2,3$. Since we need only $8$ linearly independent dyads, we omit the two other dyads that appear, which are $\ket{1-i}\bra{1+i},~i=2,3$. Consider
\begin{align}\label{eqn202}
	0=\sum_{i=1}^3\sum_{\substack{j=1\\j\ne i}}^3c_{ij}\ket{i}\bra{j}+c^\prime_1\ket{1+2}\bra{1-2}+c^\prime_2\ket{1+3}\bra{1-3}.
\end{align}
It is almost trivial to show that this is satisfied if and only if all coefficients vanish. First, take the $\bra{2}\cdots\ket{2}$ and $\bra{3}\cdots\ket{3}$ matrix elements of \myeq{eqn202}, yielding $0=c_1^\prime$ and $0=c_2^\prime$, respectively. Then, the $\bra{i}\cdots\ket{j}$ matrix element for all $j\ne i$ leads to the conclusion that $c_{ij}=0$, as well, and we are done. The chosen $8$ dyads are linearly independent, and by the symmetry between the three parties, we have verified that the conditions for Theorem~\ref{thm4} hold for the states of \myeq{eqn201}.\endproof

We can also use Theorem~\ref{thm4} to prove this set cannot be discriminated by $\locc$ even if two of the parties get together and make joint measurements on their combined (two) parts of the tripartite system. 
\begin{thm8}\label{thm8}
	The set of states in Eq.~($4$) of \cite{HalderPRL} exhibits \emph{true} strong nonlocality without entanglement.
\end{thm8}
\proof Since their combined parts have dimension equal to $d_{BC}=9$, the proof here is slightly more challenging than that for Theorem~\ref{thm7}, since we need to demonstrate there are $d_{BC}^2-1=80$ linearly independent dyads. Selecting this many dyads out of the hundreds to choose from is difficult to do by hand, but it is easy to write a short computer program that will perform this task for us. We do, indeed, find that there are $80$ linearly independent dyads satisfying the conditions of Theorem~\ref{thm4}. Given that we've already shown that for any one party, there are $d_A^2-1=8$ linearly independent such dyads, then because of symmetry between the parties, this completes the proof.\endproof 

\noindent Thus, we have shown that this set of states demonstrates what we consider to be a significantly stronger ``nonlocality" than was originally shown by the authors of Ref.~\cite{HalderPRL}.

We note that the above proof of NLWE, in the case that one views it as a tripartite system, requires only eight linearly independent dyads for each party, and it turns out that a much reduced set of states still exhibits NLWE. It is straightforward to show that the reduced set of $12$ states, $\ket{\Psi_{1\pm}},\ket{\Psi_{2\pm}},\ket{\Psi_{3\pm}}\ket{\Psi_{10\pm}},\ket{\Psi_{11\pm}},\ket{\Psi_{12\pm}}$, still exhibits nonlocality without entanglement. We have checked numerically, and it turns out that this reduced set does not exhibit (our version of) strong nonlocality without entanglement. However, omitting only the three states $\ket{i}\ket{i}\ket{i},~i=1,2,3$ does leave a strongly nonlocal set.

As another illustration of the power of Theorem~\ref{thm4}, we use it in Appendix~\ref{App2} to prove that GenTiles$1$ \cite{IBM_CMP}, a bipartite UPB on an $n\times n$ system for any even $n\ge4$, cannot be perfectly discriminated by $\locc$, a result we first obtained recently in \cite{myProdPaths}, where it was necessary to first determine the most general separable POVM that perfectly discriminates the set. Here, by using Theorem~\ref{thm4}, we are able to avoid a great deal of effort since with this approach, one need not know anything about what measurements will succeed, all one needs to know is the set of states, itself.

\subsection{Unextendible product bases consisting of the minimal number of states}\label{ssec3}
We will show in this section that a certain subset of unextendible product bases (UPB) \cite{IBM_CMP} consisting of the minimal number of states---which we will refer to as a \emph{minimal UPB}---cannot be discriminated perfectly by $\locc$. (There is a paper \cite{Rinaldis} purporting to prove that this is true for \emph{any} unextendible product basis. We believe their proof is wrong, probably in various places, and explain our reasons for this belief in Appendix~\ref{App7}.) When the UPB is on $P$ parties each having local Hilbert space $\HC_\alpha$ of dimension $d_\alpha$, the minimal number of states is given as $N=\sum_\alpha(d_\alpha-1)+1$ \cite{IBM_CMP}.

We start by showing that for a minimal UPB, every set of $d_\alpha$ of the local states making up this UPB is linearly independent.
\begin{lem2}\label{lem2}
	 A set of $N=\sum_\alpha(d_\alpha-1)+1$ pure product states on P parties is an unextendible product basis if and only if for every $\alpha$, every set of $d_\alpha$ of the local states on $\HC_\alpha$, of dimension $d_\alpha$, is linearly independent.
\end{lem2}
\noindent The proof can be found in Appendix~\ref{App5}. The following lemma will also play an important role.
\begin{lem3}\label{lem3}
	Suppose $\ket{\psi_k^{(\alpha)}}$ and linearly independent set $\{\ket{\phi_{kl}^{(\alpha)}}\}_{l=1}^{d_\alpha}$, are states on $\HC_\alpha$ of dimension $d_\alpha$. If operator $X$ is orthogonal to each of the $d_\alpha$ dyads, $\ket{\psi_k^{(\alpha)}}\bra{\phi_{kl}^{(\alpha)}}$, for fixed $k$ and $l=1,2,\ldots,d_\alpha$, then $X$ has rank strictly smaller than $d_\alpha$.
\end{lem3}
\proof Orthogonality of $X$ to each of the dyads means
\begin{align}\label{eqn302}
	\bra{\phi_{kl}^{(\alpha)}}X\ket{\psi_k^{(\alpha)}}=0,
\end{align}
for all $l$. Note that states $\ket{\phi_{kl}^{(\alpha)}}$ constitute a complete basis of $\HC_\alpha$ for each $k$, so that any state $\ket{\Phi^{(\alpha)}}\in\HC_\alpha$ can be written as a linear combination of the $\ket{\phi_{kl}^{(\alpha)}}$. Multiply \myeq{eqn302} by arbitrary complex numbers, $\mu_l$, and sum over $l$ to obtain $\bra{\Phi^{(\alpha)}}X\ket{\psi_k^{(\alpha)}}=0$. Since $\ket{\Phi^{(\alpha)}}$ is arbitrary, this means that $X\ket{\psi_k^{(\alpha)}}=0$, so $X$ cannot be full rank, and this completes the proof.\endproof

Now we are ready to prove our desired result, as codified in the following theorem.
\begin{thm2}\label{thm2}
	Given any unextendible product basis on $P$ parties, $\SC=\left\{\ket{\Psi_m}=\bigotimes_\alpha\ket{\psi_m^{(\alpha)}}\right\}$, having the minimal number of states, $N=\sum_\alpha(d_\alpha-1)+1$, with the $\alpha$th local Hilbert space $\HC_\alpha$ having dimension $d_\alpha$, if for all $\alpha$, $d_\alpha\le\sum_{\beta\ne\alpha}(d_\beta-1)+1$---equivalently, $N\ge2(d_\alpha-1)+1$---then this set of $N$ multipartite states cannot be perfectly discriminated within $\locc$.\end{thm2}
\proof According to Theorem~\ref{thm3} of the previous section, if a set of $N$ orthogonal states $\ket{\Psi_m}$ can be perfectly discriminated by $\locc$, then there exists a continuous path of product operators, $Q(s)$, starting from $I_\HC$, such that $\bra{\Psi_m}Q(s)\ket{\Psi_n}=0$ for all $m\ne n$. Restated in terms of dyads, we have that $Q(s)$ must be orthogonal to $N(N-1)=N\sum_\alpha(d_\alpha-1)$ dyads $\ket{\Psi_n}\bra{\Psi_m}$ for all $m\ne n$. Let us drop the parameter $s$ and focus on understanding the conditions under which $Q=\bigotimes_\alpha Q^{(\alpha)}$ is orthogonal to all these dyads associated with the states of a minimal UPB. More specifically, since we require a continuous path starting at $I_\HC$, there must be part of this path that consists of full-rank operators, so let us restrict to the case that $Q$, and therefore each $Q^{(\alpha)}$, are full rank.

We may partition the dyads among the parties, again as a way to distribute the job of ``being orthogonal" to $Q$. First, suppose in a given such partition, party $\alpha$ is given less than $N(d_\alpha-1)$ distinct dyads to which $Q^{(\alpha)}$ must be orthogonal. Then, there must be another party, say $\beta$, that has been given at least $N(d_\beta-1)+1$ distinct dyads to which $Q^{(\beta)}$ must be orthogonal. Each of these dyads is of the form $\ket{\psi_k^{(\beta)}}\bra{\psi_{l}^{(\beta)}}$, for some $k,l$. Since there are only $N$ distinct kets $\ket{\psi_k^{(\beta)}}$ to choose from, there must be at least one $k$ such that $\ket{\psi_k^{(\beta)}}$ is the ket appearing in $d_\beta$ of the distinct dyads. Otherwise, there is no way to account for all $N(d_\beta-1)+1$ dyads partitioned to this party. Then, by Lemma~\ref{lem2} we see that this set of bras, $\bra{\psi_l^{(\beta)}}$, spans $\HC_\beta$, so by Lemma~\ref{lem3}, $Q^{(\beta)}$ is not full rank. Therefore, since we are seeking full rank $Q$, we may  restrict to partitions that give no one party, $\alpha$, less than $N(d_\alpha-1)$ distinct dyads.

On the other hand, if any party $\alpha$ is given more than $N(d_\alpha-1)$ distinct dyads, then by the same argument just given, $Q$ cannot be full rank. Therefore, any partition allowing for full rank $Q$ must distribute exactly $N(d_\alpha-1)$ distinct dyads to party $\alpha$, for every $\alpha$.

For any such partition allowing for $Q$ to have full rank, we will next identify a set of $d_\alpha^2-1$ linearly independent dyads distributed to party $\alpha$, for all $\alpha$. This implies that for any partition consistent with full rank $Q$, there is at most one possible $Q$ orthogonal to all $d_\alpha^2-1$ dyads. Since all of these dyads are orthogonal to $I_\HC$, operators proportional to the latter are the only ones of full rank orthogonal to all these dyads and as such, satisfy the constraint that $\Pi Q\Pi$ is diagonal in the (partial) basis of the $\ket{\Psi_m}$. Therefore, there can be no \emph{continuous} path of product operators starting from $I_\HC$ and satisfying this constraint, and the proof will be complete once we demonstrate linear independence of $d_\alpha^2-1$ of the (local) dyads, for each partition and for each $\alpha$.

Note that since there are $P\ge2$ parties (and there are no UPBs on a two-qubit system), then for all $\alpha$, $N>d_\alpha+1$, so there are $N(d_\alpha-1)>(d_\alpha+1)(d_\alpha-1)=d_\alpha^2-1$ dyads distributed to each party, $\alpha$. In Appendix \ref{App3}, we show that the following set of dyads is linearly independent.
\begin{align}\label{eqn303}
	&\ket{\psi_1^{(\alpha)}}\bra{\psi_l^{(\alpha)}},~l=d_\alpha+1,\ldots,2d_\alpha-1,\notag\\
	&\ket{\psi_l^{(\alpha)}}\bra{\psi_1^{(\alpha)}},~l=d_\alpha+1,\ldots,2d_\alpha-1,\notag\\
	&\ket{\psi_k^{(\alpha)}}\bra{\phi_{kl}^{(\alpha)}},~k=2,\ldots,d_\alpha;~l=1,\ldots,d_\alpha-1,
\end{align}
where each $\ket{\phi_{kl}^{(\alpha)}}$ is one of the $\ket{\psi_m^{(\alpha)}},~m\ne k$; for each $k$, no two of the $\ket{\phi_{kl}^{(\alpha)}}$ correspond to the same $m$; and in the last line, the $\ket{\psi_k^{(\alpha)}}$ are specifically chosen to be distinct from the $\ket{\psi_l^{(\alpha)}},~l=d_\alpha+1,\ldots,2d_\alpha-1$. Note that such a set of dyads always exists for every partition and for any minimal UPB satisfying $N\ge2(d_\alpha-1)+1$: as argued above, each $\ket{\psi_m^{(\alpha)}}$ appears as the ket in $d_\alpha-1$ distinct dyads given to party $\alpha$, so any choice of $\ket{\psi_1^{(\alpha)}}$ appears with $d_\alpha-1$ of the $\ket{\psi_l^{(\alpha)}}$; and given the just-mentioned lower bound on $N$, there are more than the needed $d_\alpha-1$ \emph{other} states remaining to be chosen as the $\ket{\psi_k^{(\alpha)}},~k=2,\ldots,d_\alpha$ on the third line, each of which appear in $d_\alpha-1$ distinct dyads, which provides for the $\ket{\phi_{kl}^{(\alpha)}}$. As discussed in the preceding paragraph, this completes the proof.\endproof

\section{Conclusions}\label{sec5}
In summary, we have applied the insights of Ref.~\cite{myProdPaths} to the problem of quantum state discrimination using local operations and classical communication wherein an error is allowed but must vanish in the asymptotic limit. We obtained a lower bound on the probability of error under these circumstances and found that this lower bound provides an estimate of the correct order of magnitude relative to the known optimal error for discriminating the four Bell states. We then proved new necessary conditions that a set of mutually orthogonal states can be perfectly discriminated by $\locc$, and provided examples illustrating the power of these conditions, which greatly simplify what has previously been an extremely arduous task, that of determining whether a set of states can be discriminated with error that is vanishingly small. While for quantum state discrimination by $\locc$, the approach given in Ref.~\cite{myProdPaths} required knowledge of the precise measurement the parties were trying to implement, a key advance attained here is that they only need to know the set of states they are tasked with discriminating, nothing more.

The work we have presented here and in Ref.~\cite{myProdPaths} opens up a wide range of questions for further study. Since the numerical evaluation of our lower bound, \myeq{eqn1104}, appears to be difficult, it would be of interest for experts to develop methods of addressing this problem. At present, the greatest lower bound that we are aware of for the domino states \cite{Bennett9} is $p_{\textrm{err}}\ge1.9\times10^{-8}$ \cite{ChildsLeung}. As such, it would be interesting to know how our lower bound compares to this (perhaps surprisingly) small value.

Other avenues for further exploration include applying the ideas of Ref.~\cite{myProdPaths} to (i) strengthen those results by demonstrating the need not only for continuous paths to individual measurement outcomes, but to all outcomes of a given measurement simultaneously, which we believe we have found a way to do; (ii) find ways of determining when a quantum channel can be implemented by $\locc$, a goal we also believe we are well on the way to achieving; and finally, (iii) determine a lower bound on the error incurred when implementing a given quantum channel by $\locc$, say for example, to transform one entangled state to another.

\noindent\textit{Acknowledgments} --- We wish to thank Dan Stahlke and Jeff Kidder for helpful discussions.

\appendix
\section{Proof of Theorem~\ref{thm1}}\label{App6}
We begin by considering the error incurred by using any given POVM, $\MC_Q$ as in \eqref{eqn20}, to discriminate the set of states $\SC$, for the moment without the restriction to LOCC. Note that with $\Pi=\sum_m\sqrt{\eta_m}\Psi_m$, we have that $\Tr{\Pi^2}=\sum_m\eta_m\Tr{\Psi_m}=\sum_m\eta_m=1$, and for any complete POVM, $\{Q_i\}$, $\sum_i\Tr{Q_i\Pi^2}=\Tr{\Pi^2}=1$. In addition, $\Pi\Psi_m\Pi=\eta_m\Psi_m$. Defining $\widehat Q_i=\Pi Q_i\Pi\ge0$, we have
\begin{align}\label{eqn121}
	p_{\textrm{err}}&=1-\sum_i\max_m\eta_m\Tr{Q_i\Psi_m}=\sum_i\left[\Tr{\Pi Q_i\Pi}-\max_m\Tr{Q_i\Pi\Psi_m\Pi}\right]\notag\\
	&=\sum_i\left[\Tr{\widehat Q_i}-\max_m\Tr{\widehat Q_i\Psi_m}\right]\notag\\
	&\ge\sum_i\left[\frac{\Tr{\widehat Q_i}+\max_m\Tr{\widehat Q_i\Psi_m}}{2\Tr{\widehat Q_i}}\right]\left[\Tr{\widehat Q_i}-\max_m\Tr{\widehat Q_i\Psi_m}\right]\notag\\
	&=\sum_i\frac{1}{2\Tr{\widehat Q_i}}\left(\left[\Tr{\widehat Q_i}\right]^2-\left[\max_m\Tr{\widehat Q_i\Psi_m}\right]^2\right)\notag\\
	&\ge\sum_i\frac{1}{2\Tr{\widehat Q_i}}\left[\Tr{\widehat Q_i^2}-\sum_m\left(\Tr{\widehat Q_i\Psi_m}\right)^2\right]\notag\\
	&=\sum_i\frac{1}{2\Tr{\widehat Q_i}}\Tr{\widehat Q_i^2-2z_i\widehat Q_i+z_i^2}\notag\\
	&=\sum_i\frac{\Tr{\widehat Q_i}}{2}\normm{\frac{\widehat Q_i-z_i}{\Tr{\widehat Q_i}}}^2.
\end{align}
The third line follows from the fact that $\max_m\Tr{\widehat Q_i\Psi_m}\le\Tr{\widehat Q_i}$, while the fifth line follows from the fact that $\Tr{\widehat Q_i^2}\le\left[\Tr{\widehat Q_i}\right]^2$ for any $\widehat Q_i\ge0$, and that $0\le\left[\max_m\Tr{\widehat Q_i\Psi_m}\right]^2\le\sum_m\left[\Tr{\widehat Q_i\Psi_m}\right]^2$. In the sixth line, we introduce $z_i=\sum_m\Tr{\widehat Q_i\Psi_m}\Psi_m$, from which the equality to the preceding line follows from the fact that $\Tr{\Psi_m\Psi_n}=\delta_{mn}$. Finally in the seventh line, we introduce the definition of the Frobenius norm, which is $\normm{M}^2=\Tr{M^\dag M}=\sum_{k,l}\left|M_{kl}\right|^2$.

Let us now restrict to $\MC_Q\in$ LOCC$_\mathbb{N}$, implemented by any finite-round LOCC protocol. We will represent each such protocol as a tree graph consisting of an arbitrary number of finite branches, each branch itself consisting of a sequence of nodes. Each node $\alpha$ corresponds to a POVM element $\tilde Q_\alpha$ as described elsewhere, and each such element, once normalized to unit trace, lies at a distance $\tilde R_\alpha$ from the similarly normalized identity operator,
\begin{align}\label{eqn1101}
	\tilde R_\alpha=\normm{\frac{I_\HC}{D}-\frac{\tilde Q_\alpha}{\Tr{\tilde Q_\alpha}}},
\end{align}
with $D$ the overall dimension of $\HC$. Since it is straightforward to show that any refinement of a given measurement into rank-$1$ operators does not increase the error probability, we may assume that the outcomes of $\MC_Q$ are all rank-$1$ operators, in which case the leaf nodes, $l$, of the LOCC protocol all lie at a distance $\tilde R_l=\sqrt{(D-1)/D}$. That is, every branch of the protocol terminates at this distance from $\IC_\HC$.

We will use the following lemma to inform a truncation of LOCC protocols, see below. Define $\Delta(Q,z)=\normm{(\Pi Q\Pi-z)/\Tr{\Pi Q\Pi}}=\normm{(\widehat Q-z)/\Tr{\widehat Q}}$ and $\Delta(Q):=\min_{z\in Z_\Psi}\Delta(Q,z)$, where $z$ is an element of zonotope $\ZC_\Psi$, defined in the main text. Note for later reference that every element of $\ZC_\Psi$ is diagonal in the (partial) basis of the $\ket{\Psi_m}$. Then we have
\begin{lem1}\label{lem1}
	Consider the set $\PC_R$ of all positive semidefinite product operators acting on $\HC$ and lying at distance $R$ from the identity operator $I_\HC$, in the sense of \myeq{eqn1101}, and suppose that no operator in that set lies closer than $\Delta_R$ from zonotope $\ZC_\Psi$. That is,
	\begin{align}\label{eqn1201}
		\Delta_R=\min_{\substack{Q\in\PC_R\\z\in\ZC_\Psi}}\Delta(Q,z)=\min_{Q\in\PC_R}\Delta(Q).
	\end{align}
	Suppose, in addition, there exists node $\tilde Q_p$, parent of its child node $\tilde Q_s$, both lying along a branch of an LOCC protocol implementing measurement $\MC_Q$ and such that $\tilde R_p\le R$ and $\tilde R_s\ge R$. Then, $\Delta\left(\tilde Q_p\right)\ge\Delta_R$ or $\Delta\left(\tilde Q_s\right)\ge\Delta_R$, or both.
\end{lem1}
\proof As noted in the main text, the minimum over $z\in\ZC_\Psi$ is achieved at $z=\sum_m\Tr{\widehat Q\Psi_m}\Psi_m=\sum_m\eta_m\Tr{Q\Psi_m}\Psi_m$, for any $Q$, so in the basis of states $\ket{\Psi}$, $\widehat Q-z$ is the same as $\widehat Q$ but with its diagonal elements set to zero (note that states $\ket{\Psi_m}$ may constitute an incomplete basis here, but this is not a problem since with the use of $\Pi$ in its definition, the support of $\widehat Q$ is confined to the span of that incomplete basis). 

Consider the line segment, $Q(x)=(1-x)\tilde Q_p+x\tilde Q_s$, $0\le x\le1$, which connects $\tilde Q_p$ to $\tilde Q_s$. Since this is a continuous function of $x$, and since $\tilde R_p\le R$ and $\tilde R_s\ge R$, there exists $y$ in the range $0\le y\le1$ such that $Q(y)$ is at distance $R$ from $I_\HC$, again in the sense of \myeq{eqn1101}. $Q(y)\in\PC_R$ because we are considering an LOCC protocol, for which $\tilde Q_p$ and $\tilde Q_s$ differ only in one party's local operator. To prove the lemma, assume $\Delta(\tilde Q_p)<\Delta_R$ and $\Delta(\tilde Q_s)<\Delta_R$, which we will see leads to the condition that $\Delta(Q(y))<\Delta_R$, contradicting the definition of $\Delta_R$ as the minimum over $Q\in\PC_R$. We have,
\begin{align}\label{eqn1202}
	\Delta(Q(y))&=\min_z\normm{\frac{\widehat Q(y)-z}{\Tr{\widehat Q(y)}}}\notag\\
	&=\normm{\frac{\widehat Q(y)-z(y)}{\Tr{\widehat Q(y)}}},
\end{align}
with $\widehat Q(y)=\Pi Q(y)\Pi$ and
\begin{align}\label{eqn1203}
	z(y)&=\sum_m\Tr{\widehat Q(y)\Psi_m}\Psi_m\notag\\
	&=(1-y)\sum_m\Tr{\widehat Q_p\Psi_m}\Psi_m+y\sum_m\Tr{\widehat Q_s\Psi_m}\Psi_m\notag\\
	&=(1-y)z_p+yz_s,
\end{align}
where $\widehat Q_{p,s}=\Pi\tilde Q_{p,s}\Pi$, and $z_p,z_s$ are defined in analogy to $z(y)$. This gives
\begin{align}\label{eqn1204}
	\left[\Tr{\widehat Q(y)}\right]^2\Delta(Q(y))^2&=\normm{(1-y)(\widehat Q_p-z_p)+y(\widehat Q_s-z_s)}^2\notag\\
	&=(1-y)^2\left[\Tr{\widehat Q_p}\right]^2\Delta_p^2+y^2\left[\Tr{\widehat Q_s}\right]^2\Delta_s^2+2y(1-y)\Tr{\left[\widehat Q_p-z_p\right]\left[\widehat Q_s-z_s\right]}.
\end{align}
Noting that the inner product of unit vectors cannot exceed unity, we have that
\begin{align}\label{eqn1205}
	\Tr{\left[\widehat Q_p-z_p\right]\left[\widehat Q_s-z_s\right]}&\le\sqrt{\Tr{\left[\widehat Q_p-z_p\right]^2}\Tr{\left[\widehat Q_s-z_s\right]^2}}\notag\\
	&=\Tr{\widehat Q_p}\Tr{\widehat Q_s}\Delta_p\Delta_s,
\end{align}
and then from \myeq{eqn1204} that
\begin{align}\label{eqn1206}
	\left[\Tr{\widehat Q(y)}\right]^2\Delta(Q(y))^2\le\left[(1-y)\Tr{\widehat Q_p}\Delta_p+y\Tr{\widehat Q_s}\Delta_s\right]^2
\end{align}
By assumption, $\Delta_p<\Delta_R$ and $\Delta_s<\Delta_R$. This leads to the conclusion that
\begin{align}\label{eqn1207}
	\left[\Tr{\widehat Q(y)}\right]^2\Delta(Q(y))^2&<\left[(1-y)\Tr{\widehat Q_p}+y\Tr{\widehat Q_s}\right]^2\Delta_R^2\notag\\
	&=\left[\Tr{\widehat Q(y)}\right]^2\Delta_R^2,
\end{align}
or, $\Delta(Q(y))<\Delta_R$, a contradiction. This completes the proof.\endproof

This lemma provides a way to truncate a given finite-round LOCC protocol such that every branch that reaches a distance $R\le\sqrt{(D-1)/D}$ from $I_\HC$ is left with a (new) leaf node, $\tilde Q_\alpha$, for which $\Pi\tilde Q_\alpha\Pi$ is a distance of at least $\Delta_R$ from $\ZC_\Psi$. Recall that, since we can restrict consideration to rank-$1$ measurements, all leaf nodes lie at distance $R=\sqrt{(D-1)/D}$, and every branch corresponds to a continuous path starting at distance $R=0$. Therefore, for each branch, identify the \emph{first} node $\tilde Q_\alpha$ that is a distance at least $R$ from $I_\HC$. Truncate this branch at its parent $Q_p$, unless $\tilde Q_\alpha$ is at distance equal to $R$ or $\Delta_p<\Delta_R$, in either of which cases, truncate at $\tilde Q_\alpha$, for which the lemma tells us $\Delta_\alpha\ge\Delta_R$. 
Now we have a truncated tree for which
\begin{align}\label{eqn102}
	\Delta_\alpha=\normm{\frac{\Pi\tilde Q_\alpha\Pi-\tilde z_\alpha}{\Tr{\tilde Q_\alpha}}}\ge\Delta_R
\end{align}
for all leaf nodes, $\tilde Q_\alpha$, in the truncation, and $\tilde z_\alpha=\sum_m\eta_m\Tr{\tilde Q_\alpha\Psi_m}\Psi_m$. Each such leaf node has a set of descendants in the original protocol, which we index as $\LC_\alpha=\{l|Q_l$ is a leaf node descendant of $\tilde Q_\alpha$ in the original protocol$\}$, unless $\tilde Q_\alpha$ is itself a leaf node in the original protocol, in which case we instead define $\LC_\alpha=\{l|Q_l$ is \emph{the} leaf node $\tilde Q_\alpha$ in the original protocol$\}$. Then, $\tilde Q_\alpha=\sum_{l\in\LC_\alpha}Q_l$, and from \eqref{eqn121} we have
\begin{align}\label{eqn1103}
	p_{\textrm{err}}&\ge\sum_l\frac{\Tr{\Pi^2Q_l}}{2}\normm{\frac{\Pi Q_l\Pi-z_l}{\Tr{\Pi^2Q_l}}}^2\notag\\
	&\ge\sum_\alpha\sum_{l\in\LC_\alpha}\frac{\normm{\Pi Q_l\Pi-z_l}^2}{2\Tr{\Pi^2Q_l}}\notag\\
	&=\sum_\alpha\frac{\normm{\sum_{i\in\LC_\alpha}\left(\Pi Q_l\Pi-z_l\right)}^2}{2\sum_{i\in\LC_\alpha}\Tr{\Pi^2Q_l}}\notag\\
	&=\sum_\alpha\frac{\normm{\Pi\tilde Q_\alpha\Pi-\tilde z_\alpha}^2}{2\Tr{\Pi^2\tilde Q_\alpha}}\notag\\
	&=\sum_\alpha\frac{\Tr{\Pi^2\tilde Q_\alpha}}{2}\Delta_\alpha\notag\\
	&\ge\sum_\alpha\frac{\Tr{\Pi^2\tilde Q_\alpha}}{2}\Delta_R\notag\\
	&=\frac{1}{2}\min_{\substack{Q\in\PC_R\\z\in\ZC_\Psi}}\normm{\frac{\Pi Q\Pi-z}{\Tr{\Pi^2Q}}}^2
\end{align}
where the second (and slightly indirectly, last) lines follow since the sum over leaves descended from all of the $\tilde Q_\alpha$ includes all leaves in the original protocol; the step going from line $2$ to line $3$ is proven in Appendix~\ref{App1}; and $\tilde z_\alpha$ is defined below \myeq{eqn102}. The second-to-last line follows from Lemma~\ref{lem1}, while the last line is a result of the fact that $\tilde Q_\alpha$ is a product operator (since it is an intermediate outcome of an LOCC protocol), along with $\sum_\alpha\tilde Q_\alpha=I_\HC$.

Thus, we've derived the expression in Theorem~\ref{thm1}, which lower bounds $p_{\textrm{err}}$ for any finite number of rounds, $r$. Since this result is independent of $r$, it continues to hold in the limit $r\to\infty$, and this completes the proof.

\section{Proof of line $3$ in \myeq{eqn1103}}\label{App1}
Here we prove that
\begin{align}\label{eqnA1}
\sum_{l\in\LC_\alpha}\frac{\normm{\widehat Q_l-z_l}^2}{\Tr{\widehat Q_l}}\ge\frac{\normm{\sum_{l\in\LC_\alpha}\left(\widehat Q_l-z_l\right)}^2}{\sum_{l\in\LC_\alpha}\Tr{\widehat Q_l}},
\end{align}
with $\widehat Q_l=\Pi Q_l\Pi$. Let $t_l=\Tr{\widehat Q_l}>0$. Then, defining $M^{(l)}=\widehat Q_l-z_l$ and denoting its matrix elements in any chosen basis as $M_{\mu\nu}^{(l)}$, consider
\begin{align}\label{eqnA2}
\SC\equiv\sum_{l\in\LC_\alpha}\frac{\normm{\widehat Q_l-z_l}^2}{\Tr{\widehat Q_l}}-\frac{\normm{\sum_{l\in\LC_\alpha}\left(\widehat Q_l-z_l\right)}^2}{\sum_{l\in\LC_\alpha}\Tr{\widehat Q_l}}&=\sum_{l\in\LC_\alpha}\frac{\sum_{\mu\nu}\left|M_{\mu\nu}^{(l)}\right|^2}{t_l}-\frac{\sum_{\mu\nu}\left|\sum_{l\in\LC_\alpha}M_{\mu\nu}^{(l)}\right|^2}{\sum_{l\in\LC_\alpha}t_l}\equiv\sum_{\mu\nu}\SC_{\mu\nu}.
\end{align}
We will show that each term, $\SC_{\mu\nu}$, is non-negative. To simplify notation, let us replace $l\in\LC_\alpha$ by $l$ and take the restriction on the sums as implicit. Multiply by $\sum_jt_j>0$ and $\prod_kt_k>0$ to obtain
\begin{align}\label{eqnA3}
\SC_{\mu\nu}&=\sum_jt_j\sum_l\left|M_{\mu\nu}^{(l)}\right|^2\prod_{k\ne l}t_k-\left|\sum_lM_{\mu\nu}^{(l)}\right|^2\prod_kt_k\notag\\
	&=\sum_l\left(\prod_{k\ne l}t_k\right)\left(\sum_{j\ne l}t_j+t_l\right)\left|M_{\mu\nu}^{(l)}\right|^2-\sum_l\left(\left|M_{\mu\nu}^{(l)}\right|^2+\sum_{j\ne l}M_{\mu\nu}^{(l)\ast}M_{\mu\nu}^{(j)}\right)\prod_kt_k\notag\\
	&=\sum_l\left(\prod_{k\ne l}t_k\right)\sum_{j\ne l}t_j\left|M_{\mu\nu}^{(l)}\right|^2-\sum_l\sum_{j\ne l}M_{\mu\nu}^{(l)\ast}M_{\mu\nu}^{(j)}\prod_kt_k\notag\\
	&=\sum_l\sum_{j\ne l}\left(\prod_{k\ne l,j}t_k\right)\left[t_j^2\left|M_{\mu\nu}^{(l)}\right|^2-t_lt_jM_{\mu\nu}^{(l)\ast}M_{\mu\nu}^{(j)}\right]\notag\\
	&=\frac{1}{2}\sum_l\sum_{j\ne l}\left(\prod_{k\ne l,j}t_k\right)\left[t_j^2\left|M_{\mu\nu}^{(l)}\right|^2-t_lt_jM_{\mu\nu}^{(l)\ast}M_{\mu\nu}^{(j)}+t_l^2\left|M_{\mu\nu}^{(j)}\right|^2-t_jt_lM_{\mu\nu}^{(j)\ast}M_{\mu\nu}^{(l)}\right]\notag\\
	&=\frac{1}{2}\sum_l\sum_{j\ne l}\left(\prod_{k\ne l,j}t_k\right)\left|t_jM_{\mu\nu}^{(l)}-t_lM_{\mu\nu}^{(j)}\right|^2,
\end{align}
which is manifestly non-negative. Therefore, each $\SC_{\mu\nu}\ge0$ implying that $\SC\ge0$ as well, and this completes the proof.

\section{GenTiles$1$ cannot be perfectly discriminated by $\locc$}\label{App2}
As an illustration of how these results may be applied, we use Theorem~\ref{thm4} to prove that GenTiles$1$ \cite{IBM_CMP}, a bipartite UPB on an $n\times n$ system for any even $n\ge4$, cannot be perfectly discriminated by $\locc$, a result we first obtained recently in \cite{myProdPaths}. The states in this UPB are
\begin{align}\label{eqn1001}
	\ket{V_{km}}&=\frac{1}{\sqrt{n}}\ket{k}\otimes\sum_{j=0}^{\frac{n}{2}-1}\omega^{jm}\ket{j+k+1\Mod{n}},\notag\\
	\ket{H_{km}}&=\frac{1}{\sqrt{n}}\sum_{j=0}^{\frac{n}{2}-1}\omega^{jm}\ket{j+k\Mod{n}}\otimes\ket{k},\\
	\ket{F}&=\frac{1}{n}\sum_{ij=0}^{n-1}\ket{i}\otimes\ket{j},\notag
\end{align}
with $\omega=e^{4\pi i/n}$, $m=1,\ldots,n/2-1$ and $k=0,\ldots,n-1$. Notice that the system is symmetric under interchange of parties, so if we can show there exists a set of dyads, each of which is traceless on one party but not on the other, and which on the first party spans a subspace of dimension $n^2-1$, then this will also hold for the other party, and then by Theorem~\ref{thm4}, we will have demonstrated the desired result. The local states on the first party are
\begin{align}\label{eqn1002}
	\ket{h_{km}}&=\sum_{j=0}^{\frac{n}{2}-1}\omega^{jm}\ket{j+k\Mod{n}},\notag\\
	\ket{f}&=\sum_{j=0}^{n-1}\ket{j},
\end{align}
along with the standard basis states, $\ket{i}$.

Following a bit of guesswork and playing around numerically looking for patterns on systems with several smallish values of $n$, we have identified the following set of $n^2-1$ linearly independent, traceless dyads on the first party.
\begin{align}\label{eqn1003}
	&(1)~\ket{i}\bra{j}~~i,j=0,\ldots,n-1;~~j\ne i,i+\frac{n}{2}\mmod{n}\notag\\
	&(2)~\ket{f}\bra{h_{km}}\textrm{ and }\ket{h_{km}}\bra{f}~~k=0,m=1,\ldots,\frac{n}{2}-1\textrm{ and }k=1,m=1;\textrm{ and }\ket{f}\bra{h_{k1}}~~k=2,\ldots,\frac{n}{2}\notag\\
	&(3)~\ket{h_{km}}\bra{h_{kl}}~~k=1,m=1,l=2,\ldots,\frac{n}{2}-1\textrm{ and }k=1,m=2,l=1\notag\\
	&(4)~\ket{h_{01}}\bra{n/2}.
\end{align}
Set ($1$) corresponds to orthogonality of $\ket{H_{im}},\ket{H_{jl}}$, which are not orthogonal on the second party for at least some values of $m,l$---these are $n(n-2)$ dyads; set ($2$) corresponds to orthogonality of $\ket{F},\ket{H_{km}}$, which again, are not orthogonal on the second party---these are $2(n/2-1)+2+n/2-1=3n/2-1$; set ($3$) is for $\ket{H_{km}},\ket{H_{kl}}$ (same $k$)---$n/2-1$ dyads; and set ($4$) is for $H_{01},\ket{V_{\frac{n}{2}}m}$---which is one last dyad. The total number of these dyads is $n^2-1$, as required, and none of these are orthogonal on the second party. They are linearly independent if there is no set of non-zero coefficients satisfying the following equation.
\begin{align}\label{eqn1004}
	0&=\sum_{\substack{i,j=0\\{j\ne i,i+n/2\Mod{n}}}}c_{ij}\ket{i}\bra{j}
	+\sum_{m=1}^{n/2-1}\left(f_{0m}\ket{f}\bra{h_{0m}}+f_{0m}^\prime\ket{h_{0m}}\bra{f}\right)+f_{11}\ket{f}\bra{h_{11}}+f_{11}^\prime\ket{h_{11}}\bra{f}\notag\\
	&+\sum_{k=2}^{n/2}f_{k1}\ket{f}\bra{h_{k1}}+\sum_{m=2}^{n/2-1}g_{1m}\ket{h_{11}}\bra{h_{1m}}+g_{21}\ket{h_{12}}\bra{h_{11}}+h\ket{h_{01}}\bra{n/2}.
\end{align}
We next show that this latter equation is satisfied if and only if all its coefficients vanish, which proves that the set of dyads listed in \myeq{eqn1003} are linearly independent, spanning a space of dimension $n^2-1$. By the symmetry of the parties for GenTiles$1$, this conclusion holds for the second party, as well. Thus, by Theorem~\ref{thm4}, GenTiles$1$ cannot be perfectly discriminated within $\locc$ for any value of $n$.

We start by taking the $\bra{n/2+l}\cdots\ket{n/2+l}$ matrix elements of \myeq{eqn1003} for $l=1,\ldots,n/2-1$. Since for these values of $l$, $\inpd{n/2+l}{h_{km}}=0$ for $k=0,1$, and since the terms involving $c_{ij}$ only include non-diagonal dyads ($j\ne i$), these matrix elements of \myeq{eqn1004} yield
\begin{align}\label{eqn1005}
	0&=\sum_{k=2}^{n/2}f_{k1}\inpd{h_{k1}}{n/2+l}=\sum_{k=l+1}^{n/2}f_{k1}\omega^{k-l},
\end{align}
and we have used the facts that $\inpd{h_{k1}}{n/2+l}$ vanishes unless $k\le n/2+l\le k+n/2-1\Mod{n}$, in which case it is equal to $\omega^{k-n/2-l}$, and $\inpd{i}{f}=1$ for all $i$. Beginning with $l=n/2-1$, this reduces to $f_{n/2,1}=0$. Then, $l=n/2-2$ yields $f_{n/2-1,1}=0$, and continuing on step-by-step, reducing $l$ by unity each time, we find that $f_{k1}=0$ for all $k=2,\ldots,n/2$.

Noting that the terms involving $c_{ij}$ also do not include $j=i+n/2\Mod{n}$, we next take $\bra{n/2}\cdots\ket{0}$ to obtain
\begin{align}\label{eqn1006}
	0&=\sum_{m=1}^{n/2-1}f_{0m}+f_{11}^\prime\omega^{-1},
\end{align}
and then $\bra{n/2+l}\cdots\ket{l},~l=1,\ldots,n/2-1$ to obtain
\begin{align}\label{eqn1007}
	0&=\sum_{m=1}^{n/2-1}f_{0m}\omega^{-ml}+f_{11}\omega^{1-l}.
\end{align}
Using $\sum_{l=0}^{n/2-1}\omega^{-ml}=0$ for all $m\ne0$, we can add \myeq{eqn1006} to the sum of all versions (different $l$) of \myeq{eqn1007} to obtain $f_{11}^\prime=f_{11}\omega^2$. Similarly, from $\bra{0}\cdots\ket{n/2}$ we get
\begin{align}\label{eqn1008}
	0&=\sum_{m=1}^{n/2-1}f_{0m}^\prime+f_{11}\omega+h,
\end{align}
and from $\bra{l}\cdots\ket{n/2+l},~l=1,\ldots,n/2-1$,
\begin{align}\label{eqn1009}
	0&=\sum_{m=1}^{n/2-1}f_{0m}^\prime\omega^{ml}+f_{11}^\prime\omega^{l-1}=\sum_{m=1}^{n/2-1}f_{0m}^\prime\omega^{ml}+f_{11}\omega^{l+1}.
\end{align}
Adding \myeq{eqn1008} and all $n/2-1$ instances of \myeq{eqn1009} leaves $h=0$.

Looking now at diagonal element, $\bra{0}\cdots\ket{0}$, we have
\begin{align}\label{eqn1010}
	0&=\sum_{m=1}^{n/2-1}\left(f_{0m}+f_{0m}^\prime\right).
\end{align}
Adding \myeq{eqn1010} and all instances of \myeq{eqn1007} and \myeq{eqn1009}, we obtain
\begin{align}\label{eqn1012}
	0&=\sum_{l=0}^{n/2-1}\sum_{m=1}^{n/2-1}\left(f_{0m}\omega^{-ml}+f_{0m}^\prime\omega^{ml}\right)+\sum_{l=1}^{n/2-1}f_{11}\omega\left(\omega^l+\omega^{-l}\right).
\end{align}
which reduces to $0=-2f_{11}\omega$, so $f_{11}=0$, implying (see below \myeq{eqn1007}) $f_{11}^\prime=0$. Now \myeq{eqn1007} can be written (with $f_{11}=0$) as $M\vec f_0=\vec0$, with the elements of $\vec f_0$ being $(\vec f_0)_m=f_{0m}$, and those of $M$ are given by $M_{lm}=\omega^{-ml},~m,l\ne0$. If we add a column of all ones to obtain matrix $M^\prime$, then it is straightforward to see that $M^\prime M^{\prime\dag}$ is proportional to the $(n/2-1)$-dimensional identity operator. Thus, the rank of $M^\prime$ is $n/2-1$. However, the sum of all columns of $M^\prime$ vanishes, implying that the column rank of $M^\prime$ is the same as that of $M$. This, in turn, implies that the rank of $M$ is also $n/2-1$, so $M$ is invertible. Therefore, we have that $\vec f_0=\vec0$, and $f_{0m}=0$ for all $m=1,\ldots,n/2-1$.

Similarly, \myeq{eqn1009} can be written as $M\vec f_0^\prime=0$, with the same $M$ and $(\vec f_0^\prime)_m=f_{0m}^\prime$, and thus $f_{0m}^\prime=0$ for all $m=1,\ldots,n/2-1$, as well. Thus, the only nonzero coefficients left are the $c$'s and $g$'s.

For the $g$'s, consider all remaining diagonal elements of our constraint equations, those not previously used in the preceding arguments, $\bra{l}\cdots\ket{l},~l=1,\ldots,n/2-1$. Given our preceding results, these yield
\begin{align}\label{eqn1013}
	0&=\sum_{m=2}^{n/2-1}g_{1m}\omega^{-(m-1)l}+g_{21}\omega^{l}.
\end{align}
Defining $\vec g^T=\left(g_{21}~g_{1,n/2-1}~g_{1,n/2-2}~\ldots~g_{13}~g_{12}\right)$, this last equation may be written as $M^\ast\vec g$, with $M$ the same matrix as has appeared above. Therefore, each entry of $\vec g$ vanishes, and the only nonzero coefficients remaining are the $c_{ij}$.

That is,
\begin{align}\label{eqn1013}
	0=\sum_{\substack{i,j=0\\{j\ne i,i+n/2\Mod{n}}}}c_{ij}\ket{i}\bra{j},
\end{align}
and it is clear that $c_{ij}=0$ for all remaining $i,j$, as well. Thus, we have that \myeq{eqn1004} can be satisfied if and only if all of the coefficients appearing there vanish, showing that the set of dyads in \myeq{eqn1003} is linearly independent, which is what we set out to prove.

\section{Why Ref.~\cite{Rinaldis} is wrong}\label{App7}
Here we argue that the results of \cite{Rinaldis} purporting to prove that any unextendible product basis exhibits NLWE is wrong. It is probably wrong in various places, including the assumption that $A$ in their Eq.~($4$), the key part of which reads as
\begin{align}
	\max_{i\ne j} \left(2\lambda^2\delta^\prime\bra{\phi_i}A\ket{\phi_j}+\lambda^2\delta^{\prime2}\bra{\phi_i}A^\dag A\ket{\phi_j}\right)>\max_{i\ne j} 2\lambda^2\delta^\prime\bra{\phi_i}A\ket{\phi_j},\notag
\end{align}	
must be a positive semidefinite operator: $E=\lambda(I+\delta^\prime A)$ is positive semidefinite, but there is no apparent reason why $A$ needs to be. It is all well and good to use the polar decomposition of $S=EU$ to obtain $E\ge0$, but this implies $S=\lambda(U+\delta^\prime A^\prime)$, and there is no reason that (with the starting point being $S$) the polar decomposition of $A^\prime$ should yield the same $U$ as that for $S$. Put more simply, $E\ge0$ being close to (proportional to) the identity operator only means that the operator $A$, seen above, is small; to maintain generality, one must allow for indefinite (and even negative semidefinite) $A$. In other words, it is wrong to assume that $A$ is positive semidefinite. More importantly, the inequality in Eq.~($4$) is unjustified, even if one maximizes over the absolute value of the off-diagonal matrix element of $E^\dag E$, instead of (as it is written in their paper) maximizing over the off-diagonal matrix element itself (which may not be a real number, implying that this maximization makes no sense). The reason is that off-diagonal elements of positive operators (such as $A^\dag A$ in this equation) need not be positive (this problem arises even when considering the absolute value of the off-diagonal matrix element, as we have just suggested must be done). Indeed, these off-diagonal elements may well be complex numbers. Therefore, the inequality in their Eq.~($4$) could very easily be in the opposite direction, thus failing to provide the \emph{critical} lower bound on the degree to which the states remain orthogonal.

The argument given in Ref.~\cite{Rinaldis} has been criticized previously \cite{KKB} for a different technical point, a criticism that we do not understand. We do take issue with that same technical point, however, but for different reasons. Below Eq.~($16$) of Ref.~\cite{Rinaldis}, for the case $c_N\to0$, the claim that Eqs.~($5$) and ($6$) lead to $\epsilon\to0$ appears to have no basis. It is not entirely clear how they arrive at this conclusion, but they seem to be assuming that if the off-diagonal elements of $O(N)$ all vanish, then its diagonal elements must all be equal to each other, an assumption for which we see no justification. In any case, if their Eq.~($4$) is wrong, as we have suggested above, then it is almost certain that the analogous Eq.~($15$) is also wrong, calling into question the validity of the critical Eq.~($16$).

Finally, given the fact that there are several apparent erroneous assumptions made in the proof---along with numerous apparent typographical errors and omissions of detailed explanations of their reasoning---one is left with little confidence in their conclusions, even if one could follow the proof in detail, which is not an easy task.

\section{Proof of Lemma~\ref{lem2}}\label{App5}
\proof We prove the ``only if" direction by way of contradiction. Thus, suppose we have a minimal UPB and there exists party $\alpha$ and a linearly \emph{dependent} set of $d_\alpha$ local states, say $\{\ket{\psi_m^{(\alpha)}}\}_{m=1}^{d_\alpha}$. Then, there exists a partition of the entire set of $N$ states such that $\{\ket{\Psi_m}\}_{m=1}^{d_\alpha}$ are placed with party $\alpha$, and all the other parties, $\beta$, are each given $d_\beta-1$ states. Note that for every party, the local states corresponding to this partition fail to span the entire local Hilbert space. Therefore, for each party $\beta$, we can identify one additional local state, $\ket{\phi^{(\beta)}}$ orthogonal to all the $d_\beta-1$ local states apportioned to party $\beta$ for $\beta\ne\alpha$, and since the $d_\alpha$ local states partitioned to party $\alpha$ also do not span the local Hilbert space, we can do the same for party $\alpha$ with state $\ket{\phi^{(\alpha)}}$. Taking the tensor product of these $P$ local states, we obtain a product state $\bigotimes_\mu\ket{\phi^{(\mu)}}$, orthogonal to all $N$ of the original states of the UPB, extending the UPB and contradicting the fact that it is a UPB to begin with. This completes the proof of the only if part.

To prove the other direction, simply notice that for every partition of the states among the parties, there is at least one party that is given at least $d_\alpha$ local states. By assumption, the set of states given to that party is linearly independent, spanning $\HC_\alpha$, and thus, there can be no state on $\HC_\alpha$ orthogonal to those $d_\alpha$ local states. Therefore, one cannot extend the original set of states by adding one more orthogonal product state. That is, the set is unextendible.\endproof

\section{Linear independence of dyads for Theorem~\ref{thm2}}\label{App3}
Here, we show that the set of dyads in \myeq{eqn303} is linearly independent. To this end, consider
\begin{align}\label{eqn304}
	0=\sum_{l=1}^{d_\alpha-1}\left(\sum_{k=2}^{d_\alpha}c_{kl}\ket{\psi_k^{(\alpha)}}\bra{\phi_{kl}^{(\alpha)}} +
	c_{1l}\ket{\psi_1^{(\alpha)}}\bra{\psi_{d_\alpha+l}^{(\alpha)}} +
	c_{d_\alpha+1,l}\ket{\psi_{d_\alpha+l}^{(\alpha)}}\bra{\psi_1^{(\alpha)}}\right).	
\end{align}
Since according to Lemma~\ref{lem2}, $\ket{\psi_k^{(\alpha)}},~k=1,\ldots,d_\alpha$ are a basis of $\HC_{\alpha}$, we may expand
\begin{align}\label{eqn305}
	\ket{\psi_{d_\alpha+l}^{(\alpha)}}=\sum_{k=1}^{d_\alpha}\mu_{kl}\ket{\psi_k^{(\alpha)}},
\end{align}
and then \myeq{eqn304} becomes
\begin{align}\label{eqn306}
	0=\sum_{l=1}^{d_\alpha-1}\left[\sum_{k=2}^{d_\alpha}\ket{\psi_k^{(\alpha)}}\left(c_{kl}\bra{\phi_{kl}^{(\alpha)}} + c_{d_\alpha+1,l}\mu_{kl}\bra{\psi_1^{(\alpha)}}\right) +
	\ket{\psi_1^{(\alpha)}}\left(\sum_{k=1}^{d_\alpha}c_{1l}\mu_{kl}^\ast\bra{\psi_k^{(\alpha)}}+c_{d_\alpha+1,l}\mu_{1l}\bra{\psi_{1}^{(\alpha)}}\right)\right].
\end{align}
Since $\ket{\psi_k^{(\alpha)}},~k=1,\ldots,d_\alpha$ are linearly independent, each of their coefficients in the preceding equation must vanish separately. That is,
\begin{align}\label{eqn307}
	0=\sum_{l=1}^{d_\alpha-1}\left(c_{kl}\bra{\phi_{kl}^{(\alpha)}} + c_{d_\alpha+1,l}\mu_{kl}\bra{\psi_1^{(\alpha)}}\right),~k=2,\ldots,d_\alpha
\end{align}
and
\begin{align}\label{eqn308}
	0=\sum_{l=1}^{d_\alpha-1}\left(\sum_{k=1}^{d_\alpha}c_{1l}\mu_{kl}^\ast\bra{\psi_k^{(\alpha)}}+c_{d_\alpha+1,l}\mu_{1l}\bra{\psi_{1}^{(\alpha)}}\right)=\sum_{l=1}^{d_\alpha-1}\left[\sum_{k=2}^{d_\alpha}c_{1l}\mu_{kl}^\ast\bra{\psi_k^{(\alpha)}}+(c_{d_\alpha+1,l}\mu_{1l}+c_{1l}\mu_{1l}^\ast)\bra{\psi_{1}^{(\alpha)}}\right].
\end{align}
Noting that the set of $d_\alpha$ states $\{\bra{\psi_1^{(\alpha)}},\{\bra{\phi_{kl}^{(\alpha)}}\}_{l=1}^{d_\alpha-1}\}$ is also linearly independent for each $k$, the coefficients of these states must each vanish in \myeq{eqn307}, implying $c_{kl}=0$ for all $k=2,\ldots,d_\alpha$ and all $l=1,\ldots,d_\alpha-1$. In addition,
\begin{align}\label{eqn309}
	\sum_{l=1}^{d_\alpha-1}c_{d_\alpha+1,l}\mu_{kl}=0,~k=2,\ldots,d_\alpha.
\end{align}
Considering \myeq{eqn305}, we find
\begin{align}\label{eqn310}
	\sum_{l=1}^{d_\alpha-1}c_{d_\alpha+1,l}\ket{\psi_{d_\alpha+l}^{(\alpha)}}=\sum_{k=1}^{d_\alpha}\sum_{l=1}^{d_\alpha-1}c_{d_\alpha+1,l}\mu_{kl}\ket{\psi_{k}^{(\alpha)}}=\left(\sum_{l=1}^{d_\alpha-1}c_{d_\alpha+1,l}\mu_{1l}\right)\ket{\psi_1^{(\alpha)}},
\end{align}
and we have used \myeq{eqn309}. This implies linear dependence of the $d_\alpha$ states $\ket{\psi_1^{(\alpha)}}$ and $\ket{\psi_{d_\alpha+l}^{(\alpha)}},~l=1,\ldots,d_\alpha-1$, a contradiction, unless $c_{d_\alpha+1,l}$ vanishes for all $l$. These results reduce \myeq{eqn308} to
\begin{align}\label{eqn311}
	0=\sum_{l=1}^{d_\alpha-1}c_{1l}\left(\sum_{k=2}^{d_\alpha}\mu_{kl}^\ast\bra{\psi_k^{(\alpha)}}+\mu_{1l}^\ast\bra{\psi_1^{(\alpha)}}\right).
\end{align}
Once again, the coefficient of each $\ket{\psi_k^{(\alpha)}}$ must vanish separately, implying $\sum_lc_{1l}\mu_{kl}=0$ for all $k$. Similarly to what we just did in \myeq{eqn310}, consider
\begin{align}\label{eqn312}
	\sum_{l=1}^{d_\alpha-1}c_{1l}^\ast\ket{\psi_{d_\alpha+l}^{(\alpha)}}=\sum_{k=1}^{d_\alpha}\sum_{l=1}^{d_\alpha-1}c_{1l}^\ast\mu_{kl}\ket{\psi_{k}^{(\alpha)}}=0,
\end{align}
and we have used \myeq{eqn311}. This implies that the set of $d_\alpha-1$ states in the sum on the left is linearly dependent, which contradicts Lemma~\ref{lem2} unless $c_{1l}=0$ for all $l$.

Collecting all these results, we see that the sum of $d_\alpha^2-1$ dyads in \myeq{eqn304} vanishes if and only if each of the $c_{mn}$ appearing there vanishes identically. That is, those $d_\alpha^2-1$ dyads constitute a linearly independent set. As argued in the paragraph preceding \myeq{eqn303}, this completes the proof.



%

\end{document}